\documentclass[epjST]{svjour}
\usepackage{graphicx}
\usepackage{amsmath}
\usepackage{amssymb}
\begin{document}
\title{Forced synchronization of an oscillator with a line of equilibria}

\author{Ivan A. Korneev\inst{1} \and Andrei V. Slepnev\inst{1} \and Vladimir V. Semenov\inst{2}\thanks{\email{semenov.v.v.ssu@gmail.com}} \and Tatiana Vadivasova\inst{1}}
\institute{Department of Physics, Saratov State University, Astrakhanskaya str. 83, 410012 Saratov, Russia \and FEMTO-ST Institute/Optics Department, CNRS \& University Bourgogne Franche-Comt\'e, \\15B avenue des Montboucons,
Besan\c con Cedex, 25030, France}
\abstract{
The model of a non-autonomous memristor-based oscillator with a line of equilibria is studied. A numerical simulation of the system driven by a periodical force is combined with a theoretical analysis by means of the quasi-harmonic reduction. Both two mechanisms of synchronization are demonstrated: capture of the phase and frequency of oscillations and suppression by an external signal. Classification of undamped oscillations in an autonomous system with a line of equilibria as a special kind of self-sustained oscillations is concluded due to the possibility to observe the effect of frequency-phase locking in the same system in the presence of an external influence. It is established that the occurrence of phase locking in the considered system continuously depends both on parameter values and initial conditions. The simultaneous dependence of synchronization area boundaries on the initial conditions and the parameter values is also shown. 
} %end of abstract
\maketitle

\section{Introduction}

The effect of synchronization first described in 1665 by C. Huygens \cite{hugenii1986} represents a fundamental property of dynamical systems \cite{pikovsky2001,mosekilde2002,balanov2009}. The synchronization associated with phase locking and capture of the natural spectral line (the synchronization in the sense of Huygens) is a significant feature of self-oscillatory systems. Depending on specifics of explored objects the synchronization can be exhibited in a different manner. Thus, stochastic synchronization \cite{neiman1994,shulgin1995,han1999,lindner2004,goldobin2005}, synchronization of chaos \cite{rosenblum1996,anishchenko2007}, lag synchronization \cite{rosenblum1997}, generalized synchronization \cite{abarbanel1996,hramov2005} are distinguished besides the classical forms of mutual and forced synchronization. The phenomenon of synchronization is not limited to nonlinear dissipative systems with a finite number of attractors, but also applies to Hamiltonian systems \cite{heagy1994,hampton1999,hannachi1999}.

Despite synchronization is well-studied for different kinds of dynamical systems, there exists a class of dissipative dynamical systems for which this effect has not been researched in terms of the Huygens treatment of phase-frequency locking. Such systems are oscillators with m-dimensional manifolds of equilibria which consist of non-isolated equilibrium points. In the simplest case these manifolds exist as a line of equilibria (m = 1). Normally hyperbolic manifolds of equilibria are distinguished and their equilibria are characterized by m purely imagine or zero eigenvalues, whereas all the other eigenvalues have nonzero real parts. In terms of the dynamical system theory, the systems whose phase space includes normally hyperbolic manifolds of equilibria can be referred to a special kind of systems with unusual characteristics. Particularly, it has been shown that their significant feature is the occurrence of bifurcations without parameters \cite{fiedler2000-1,fiedler2000-2,fiedler2000-3,liebscher2015,riaza2012,riaza2018}, i.e., the bifurcations corresponding to fixed parameters when the condition of normal hyperbolicity is violated at some points of the manifold of equilibria. 

Bifurcation transitions in systems with a line of equilibria can give rise to the appearance of periodic solutions with the undamped amplitude. In such a case all the fixed points from some part of the line of equilibria bifurcate simultaneously and a manifold of non-isolated closed curves appears in their vicinity.  After that the system has an attractor including a continuous set of the closed curves. The described transformations have been observed in models of memristor-based electronic circuits \cite{itoh2008,messias2010,botta2011,semenov2015}, where the existence of the line of equilibria is associated with specifics of a memristor state equation \cite{chua1971,chua1976}. Excitation of periodic oscillations in models of circuits including the memristor has been theoretically analyzed by means of the quasi-harmonic reduction \cite{korneev2017,korneev2017-2}. It has been established that the oscillation excitation has distinctive features of the Andronov-Hopf bifurcation. This conclusion allows to draw an analogy between classical self-oscillators with a finite number of steady states and limit cycles and dynamical systems with a line of equilibria. However the principal difference takes place. The invariant closed curve is not an attractor in itself. Despite each closed curve corresponds to undamped periodical oscillations in a dissipative system, it does not satisfy the definition of the Andronov-Poincar\'{e} limit cycle. The invariant closed curve has no its own limited basin of attraction. Then the question on whether periodical oscillations in an autonomous system with a line of equilibria can be classified as self-sustained oscillations is raised. On the one hand, the answer is 'Yes'. If a system is autonomous, dissipative, nonlinear and demonstrates undamped oscillations, then it represents a self-oscillator. On the other hand, an image of the periodic self-oscillations is a stable limit cycle, which is absent in the phase space of the system with a line of equilibria. In this situation  the self-oscillatory character of the dynamics can be confirmed by observation of the phase-frequency synchronization \cite{pikovsky2001}.

In the present paper we consider the model of a memristor-based quasi-harmonic oscillator driven by an external periodical influence. The nonlinear conductivity of the memristor is responsible for limitation of amplitude growth and the existence of a line of equilibria. The autonomous dynamics of the studied system has been described in details \cite{semenov2015,korneev2017,korneev2017-2}. The further step is consideration of the non-autonomous system. The current research is inspired by two purposes. The first one is the issue of classification of periodical undamped oscillations in dynamical systems with a line of equilibria. If the existence of phase-frequency locking is manifested in the forced system with a line of equilibria, then its undamped periodical oscillations can be called 'self-oscillations'. The second aim is to reveal intrinsic peculiarities of the synchronization caused by a continuous dependence of oscillation characteristics on an initial state being typical for an autonomous system with a line of equilibria. 

\section{Model and methods}

\begin{figure}[t!]
\centering
\parbox[c]{.4\linewidth}{
\includegraphics[width=\linewidth]{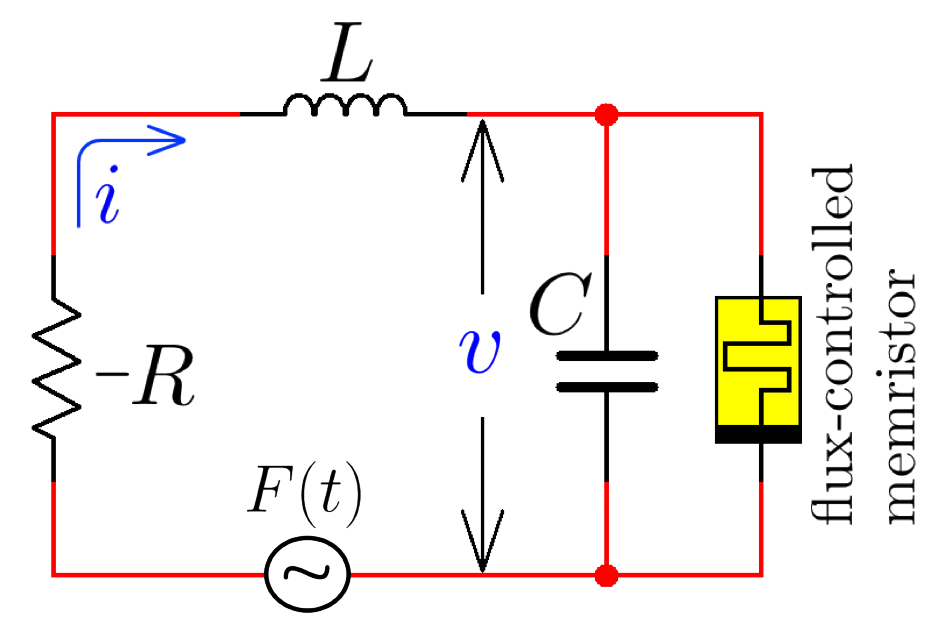}
}
\hspace{1cm}
\parbox[c]{.33\linewidth}{
\includegraphics[width=\linewidth]{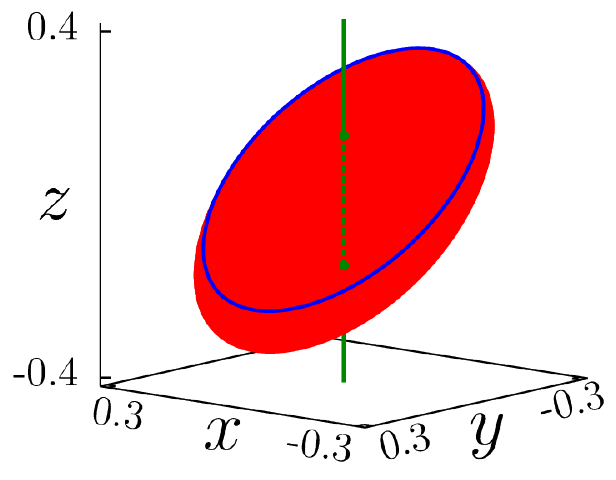}
}
\center (a)   \hspace{7cm}    (b)
\caption{(a) Schematic circuit diagram of model (\ref{model}); (b) Attractor of autonomous (at $F(t)\equiv 0$) system (\ref{model}) for parameter values $\alpha=\gamma=1$, $\beta = 0.035$, $a = 0.02$, $b = 0.8$. The red surface marks a continuous set of invariant closed curves (one of them is coloured in blue), the green solid lines are attractive manifolds of a line of equilibria, the dashed green line is a repelling manifold of the line of equilibria.}
\label{fig1}
\end{figure}

Figure \ref{fig1} (a) shows a series-oscillatory circuit including the capacitor $C$, the inductor $L$, the constant negative resistance $-R$, and a voltage source $F(t)$. The system also contains a flux-controlled memristor  parallel-connected to the capacitor. An autonomous modification without $F(t)$ of the depicted circuit is described in \cite{semenov2015,korneev2017,korneev2017-2} in details. In the current paper the non-autonomous system in Fig. \ref{fig1} is considered in the following dimensionless form:
\begin{eqnarray}
\label{model}
\dot{x}= \alpha (y - G_{M}(z)x), ~~~
\dot{y}= -\gamma x + \beta y + F(t), ~~~
\dot{z} = x, 
\end{eqnarray}
where $x \sim v$ and $y \sim i$ are dynamical variables being proportional to the the voltage on the capacitor $C$ and the current through the inductor $L$ correspondingly. The dynamical variable $z$ corresponds to the flux controlling the memristor (the quantity 'flux' is introduced as the integral $\int^{t}_{0}vdt$, see \cite{chua1971,chua1976}).  Dimensionless parameters $\alpha$, $\beta$ and $\gamma$ ($\alpha \sim 1/C$, $\beta \sim R/L $ and $\gamma \sim 1/L$) are assumed to be positive. In order to simplify the system, the parameters $\gamma$ and $\alpha$ are set to be equal to unity. The term $F(t)$ is the harmonic external impact, $F(t)=B \cos {(\omega_{ex} t)}$. A function $G_M(z)$ describes the memristor conductivity. The function $G_{M}(z)$ corresponding to the cubic memristor model \cite{messias2010,korneev2017-2,chua2011,liu2015} is used:
\begin{eqnarray}
\label{GM}
G_{M} = a + bz^{2},
\end{eqnarray}
where $a$ and $b$ are memristor parameters: $a=0.02$, $b=0.8$. 
%
%%%%%%%%%%%%%%%%%%%%%%%%%%% Fig.1 %%%%%%%%%%%%%%%%%%%%%%%%%%%%%%%%%
%
%%%%%%%%%%%%%%%%%%%%%%%%%%%%%%%%%%%%%%%%%%%%%%%%%%%%%%%%%%%%%%%%%%

The autonomous (at $B = 0$) system (\ref{model}) with the nonlinearity (\ref{GM}) has a line of equilibria in the phase space. Indeed, each point with the coordinates ($x=y=0$, $z\in(-\infty, \infty)$) is an equilibrium point. One of the eigenvalues of the equilibria is always equal to zero and the others depend on the parameters and the position of the point on the line of equilibria. In case $\beta \in (0,a)$, an attractor of the system is a manifold of equilibria and all trajectories are attracted to him. Growth of the parameter $\beta$ gives rise to bifurcation change at $\beta=a$. The part of the line of equilibria corresponding to the points with the coordinates ($x=y=0$, $|z|<\sqrt{(\beta-a)/b}$) becomes unstable. A manifold of invariant closed curves appears in the vicinity of equilibria ($x=y=0$, $|z|<\sqrt{(\beta-a)/b}$). Then the system attractor consists of a continuous set of closed curves (one of them is coloured in blue in Fig. \ref{fig1} (b)) forming a two-dimensional surface (red surface in Fig. \ref{fig1} (b)) and attractive sets of the line of equilibria (solid green lines in Fig. \ref{fig1} (b)). Depending on the initial conditions, trajectories in the phase space are attracted to one of the closed curves or to a point on the $OZ$ axis. Thus, one can obtain various oscillatory regimes at fixed parameter values and different initial conditions: either motion to an equilibrium point or periodic oscillations with different undamped amplitudes.

Further consideration of the non-autonomous system (\ref{model}) is focused on the dynamics at $\beta>a$. The system is studied by means of a numerical simulation and a theoretical analysis. Numerical modelling is carried out by using the fourth-order Runge-Kutta method with the time step $\Delta t =0.001$. Theoretical approach implies search of a solution of the system (\ref{model}) in the frameworks of the quasi-harmonic reduction.

\section{Quasi-harmonic analysis}

Let us consider non-autonomous oscillations of the memristive oscillator (\ref{model}) with the nonlinearity (\ref{GM}) at the frequency of the external impact, $\omega_{ex}$, close to harmonic ones:
\begin{eqnarray}
\label{zamena}
x(t) &=& \frac{1}{2} \left( A(t)e^{j\omega_{ex}t}+ A^{\ast}(t)e^{-j\omega_{ex}t} \right), \nonumber \\
\dot{x}(t) &=& \frac{j\omega_{ex}}{2} \left( A(t)e^{j\omega_{ex}t}- A^{\ast}(t)e^{-j\omega_{ex}t} \right),\\
z(t) &=& z_{0}-\frac{j}{2\omega_{ex}} \left( A(t)e^{j\omega_{ex}t}- A^{\ast}(t)e^{-j\omega_{ex}t} \right), \nonumber
\end{eqnarray} 
where $A(t) = \rho (t) e^{j \varphi (t)}$ is the complex amplitude of oscillations, which is assumed to be a slowly varying function during the oscillation period $T_{ex} = 2 \pi / \omega_{ex}$, $z_{0} = z(0)$ is the initial value of the variable $z$, $j = \sqrt{-1}$ is the imaginary unit, the symbol '$\ast$' means the complex conjugation. The expression for the variable $z(t)$ corresponds to the assumption $x(0) = 0$. Applying the averaging method described in details in \cite{korneev2017-2} for the case of the autonomous memristive oscillator (\ref{model}), one obtains the following averaged equation for the complex amplitude:
\begin{eqnarray}
\label{eq-A}
\dot{A} = \frac{A}{2} \left( \beta - a -\frac{b|A|^{2}}{4\omega^{2}_{ex}} -bz^{2}_{0} \right)- \frac{jA}{2} \left[2\Delta + \frac{b\beta}{\omega_{ex}} \left(z^{2}_{0}+\frac{|A|^{2}}{4\omega^{2}_{ex}} \right) \right]-\frac{jB}{2\omega_{ex}},
\end{eqnarray}
where the detuning parameter $\Delta$ is introduced as
\begin{eqnarray}
\label{Delta}
\Delta=\frac{\omega^{2}_{ex}-\omega^{2}_{0}}{2\omega^{2}_{ex}}\approx \omega_{ex}-\omega_{0},
\end{eqnarray}
$\omega_{0} = 1- \beta a$ is the natural frequency of oscillations in the absence of the external influence.

If we set $B = 0$, $\Delta = 0$ and replace $\omega_{ex}$ with $\omega_{0}$, then the equation (\ref{eq-A}) fully coincides with the corresponding equation for the complex amplitudes of the memristive system (\ref{model}) in the autonomous case obtained in \cite{korneev2017-2}.

Separating the real and imaginary parts of the equation (\ref{eq-A}), we obtain a system of two equations for the real amplitude $\rho$ and phase $\varphi$:
\begin{eqnarray}
\label{eq-rho-phi}
\dot{\rho} &=& \frac{\rho}{2} \left( \beta - a - bz^{2}_{0} - \frac{b \rho^{2}}{4\omega^{2}_{ex}} \right) - \frac{B}{2\omega_{ex}}\sin{\varphi}, \nonumber \\
\dot{\varphi} &=& -\Delta - \frac{b \beta}{2\omega_{ex}} \left( z^{2}_{0} + \frac{\rho^{2}}{4\omega^{2}_{ex}} \right) - \frac{B}{2\rho\omega_{ex}}\cos{\varphi}.
\end{eqnarray}
Here $\varphi$ is the instantaneous phase difference between the external impact $F(t)$ and the oscillations of the memristive oscillator. It can be seen from Eqs. (\ref{eq-rho-phi}) that the initial value of the variable $z$ (the quantity $z_{0}$) affects not only the amplitude of oscillations, but also the dynamics of the phase difference between the oscillations and the external force. One can introduce an effective detuning value:
\begin{eqnarray}
\label{Delta-eff}
\Delta_{eff} = -\Delta - \dfrac{b\beta}{2\omega_{ex}}\left( z_{0}^2+\frac{\rho^2}{4\omega_{ex}^2} \right),
\end{eqnarray}
which depends both on $z_{0}$ and $\rho$. We assume that the frequency of impact $\omega_{ex}$ is close to the natural frequency $\omega_{0} \approx 1$ and the amplitude of oscillations $\rho$ differs little from the amplitude of unperturbed oscillations $\rho_{0} $ for the same initial conditions, which depends not only on the parameters, but also on the initial value $z_{0}$:
\begin{eqnarray}
\label{rho-0}
\rho_{0} = 2\sqrt{\frac{\beta - a}{b} - z^{2}_{0}}.
\end{eqnarray}
Using the assumptions mentioned above, one can approximately describe the phenomenon of forced synchronization only by the equation of the phase $\varphi$ dynamics:
\begin{eqnarray}
\label{eq-phi}
\dot{\varphi}= \Delta_{eff} - \frac{B}{2\rho_{0}}\cos{\varphi},
\end{eqnarray}
where $\Delta_{eff} = -\Delta - \frac{\beta}{2}(\beta - a)$.
Equation (\ref{eq-phi}) is the simplest model of phase synchronization known as the Adler equation. It implies the condition of phase capture, when the phase difference $\varphi$ between the oscillations of the driven system and the external impact is constant:
\begin{eqnarray}
\label{cond}
|\Delta_{eff}| \leq \frac{B}{2\rho_{0}}.
\end{eqnarray}
After substitution of the value $\rho_{0}$ instead of $\rho$ into the phase equation (\ref{eq-rho-phi}), the dependence of the effective detuning (see the formula (\ref{Delta-eff})) on the initial value $z_ {0}$ disappears. At the same time the reduced amplitude of the impact $\frac{B}{2 \rho_{0}}$ in the condition (\ref{cond}) depends on the value $z_{0}$. This circumstance leads to the possibility to control the effect of phase locking by choosing the initial condition $z_{0}$. The boundary of the phase-locking region corresponding to the condition (\ref{cond}) is continuously shifted within certain limits of the variation of $z_{0}$.

In the presence of an impact, the average frequency of oscillations $\tilde{\omega}_{0}$ differs from the unperturbed value $\omega_{0}$. One can introduce the average difference frequency (beat frequency) as being:
\begin{eqnarray}
\label{Omega-0}
\Omega=<\dot{\varphi}> = <\tilde{\omega}_{0} (t)>-\omega_{ex} = \lim_{t \rightarrow \infty} {\frac{\varphi(t)-\varphi(0)}{t}},
\end{eqnarray}
which is equal to zero in the synchronization region, since the frequency of oscillations is equal to the frequency of the impact. The brackets $<\ldots>$ mean time averaging. For the phase model (\ref{eq-phi}), the beat frequency is determined by the approximate expression
\begin{eqnarray}
\label{Omega-1}
\Omega=\sqrt{ \Delta^{2}_{eff}-\left(\frac{B}{2\rho_{0}}\right)^{2}}.
\end{eqnarray}
With allowance for (\ref{Delta}) and (\ref{Delta-eff}), one can obtain the dependence of the beat frequency $\Omega$ on the frequency of exposure $\omega_{ex}$:
\begin{eqnarray}
\label{Omega-2}
\Omega=\sqrt{ \left( \frac{\omega^{2}_{ex}-(1-\beta a)^{2}}{2\omega^{2}_{ex}} + \frac{\beta}{2}(\beta - a) \right)^{2}-\left(\frac{B}{2\rho_{0}}\right)^{2}}.
\end{eqnarray}
Formula (\ref{Omega-2}) also implies a simultaneous dependence both on parameter values and initial conditions as well as the condition (\ref{cond}). Consequences of such dependence are considered in the further section, where results of a numerical simulation and analytical solution are compared.

\section{Numerical experiment}

During numerical study of synchronization, the instantaneous phase of oscillations was determined from results of integration of the equations (\ref{model}) as being
\begin{eqnarray}
\label{Phase-t}
\Phi(t)=\arctan{\left( \frac{y(t)}{x(t)}\right)} \pm \pi k,
\end{eqnarray}
where $k$ is an integer variable, which value at each time is determined by the requirement of the function $\Phi(t)$ continuity. Using the phase difference $\varphi(t) = \Phi(t) - \omega_{ex} t$, it is easy to calculate the average beat frequency (\ref{Omega-1}). At small values of the impact amplitude, one can expect a fairly good agreement between the results obtained for the original equations (\ref{model}) and the approximate model (\ref{eq-phi}).

\begin{figure}[h!]
\centering
\hspace{0.1cm}
\parbox[c]{.45\linewidth}{
  \includegraphics[width=\linewidth]{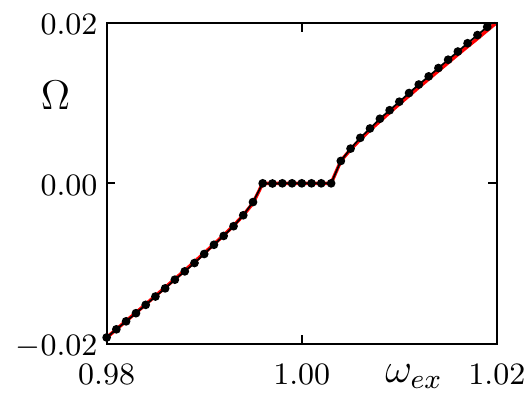}
}
\hspace{0.2cm}
\parbox[c]{.45\linewidth}{
  \includegraphics[width=\linewidth]{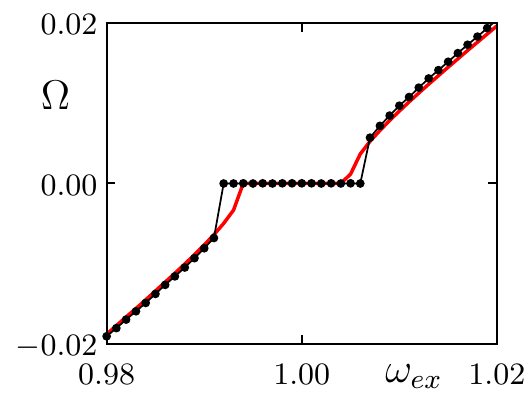}
}
\vspace{-2mm}
\center \hspace{1cm}  (a)   \hspace{7cm}    (b)

\centering
\parbox[c]{.47\linewidth}{
  \includegraphics[width=\linewidth]{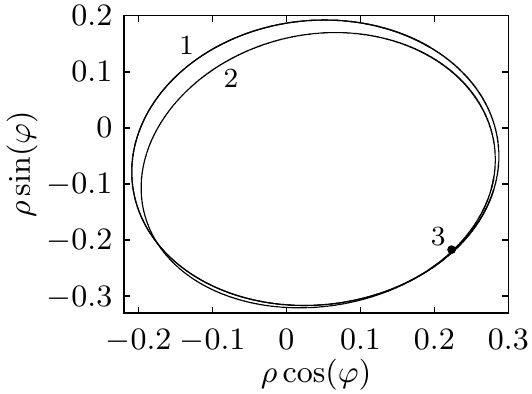}
}
\parbox[c]{.47\linewidth}{
  \includegraphics[width=\linewidth]{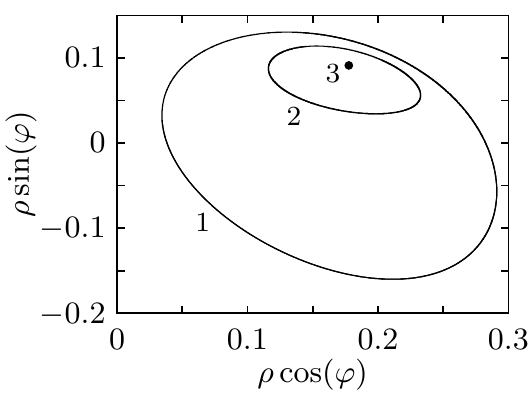}
}
\vspace{-2mm}
\center \hspace{1cm}  (c)   \hspace{7cm}    (d)

\caption{Forced synchronization of memristive oscillator for fixed initial conditions. (a),(b) Dependence of  average beat frequency $\Omega$ on external influence frequency $\omega_{ex}$ obtained numerically for Eqs. (\ref{model}) (black circles) and using approximate expressions (\ref{Omega-2}) (red solid line) at $B = 0.002$ (panel (a)) and $B = 0.003$ (panel (b)); (c),(d) Phase portraits of system (\ref{eq-rho-phi}) at $B = 0.002$ (panel (c)) and $B = 0.003$ (panel (d)). Chosen values $\omega_{ex}$ for the panel (c) are: $\omega_{ex} = 0.993$ (curve 1), $0.995$ (curve 2), $0.997$ (point 3). Chosen values $\omega_{ex}$ for the panel (d) are: $\omega_{ex} = 0.992$ (curve 1), $0.9925$ (curve 2), $0.9928$ (point 3). Initial conditions for the source system (\ref{model}): $x(0) = y(0) = z(0) = 0$. Initial conditions for the truncated equations (\ref{eq-rho-phi}): $ \rho(0) = 0.0001,~\varphi(0) = 0,z(0) = 0$. Other parameters are: $\alpha = 1$, $\beta = 0.035$, $\gamma = 1$, $a = 0.02$, $b = 0.8$.}
\label{fig2}
\end{figure}

Calculations of the dependence of the average beat frequency $\Omega$ on the frequency of external influence $\omega_{ex}$ for the initial condition $x(0) = y(0) = z(0) = 0$ have been obtained both numerically for the equations (\ref{model}) and analytically using the approximate expression (\ref{Omega-2}). The results for two values of the impact amplitude are shown in Fig. \ref{fig2} (a),(b). When the amplitude of the impact is $B = 0.002$, there is a good agreement between the numerical result and the approximate estimation (\ref{Omega-2}) [Fig. \ref{fig2} (a)]. At the same time, at $B = 0.003$ some differences in the width of the synchronization region are noticeable. In addition, the synchronization mechanism changes [Fig. \ref{fig2} (b)]. Obtained in the numerical simulation of the system (\ref{model}) sharp jump in the beat frequency at the boundary of the synchronization region shows that at $B = 0.003$ the synchronization mechanism is oscillation suppression, which cannot be described using the phase model (\ref{eq-phi}).
 
%
%%%%%%%%%%%%%%%%%%%%%%%%%%% Fig.2 %%%%%%%%%%%%%%%%%%%%%%%%%%%%%%%%%%%%%%%%%%%%%%%%%%%%
%
%%%%%%%%%%%%%%%%%%%%%%%%%%%%%%%%%%%%%%%%%%%%%%%%%%%%%%%%%%%%%%%%%%%%%%%%%%%%%%%%%%%%%%%%
%
Differences in the character of transition to the synchronization at $B = 0.002$ and $B = 0.003$ obtained using the system of truncated equations (\ref{eq-rho-phi}) are illustrated in Fig. \ref{fig2} (c),(d). Since there is a division by $\rho $ in the right part of the phase equation (\ref{eq-rho-phi}), an integration of the system (\ref{eq-rho-phi}) was carried out from initial conditions being somewhat different from zero: $\rho(0) = 0.0001,~\varphi(0) = 0, z(0) = 0$. Smallness of the value $\rho(0)$ provides access to almost the same steady state as zero initial conditions in the system (\ref{model}).  Phase portraits  of the system (\ref{eq-rho-phi}) on the plane  ($\rho \cos{\varphi},~\rho \sin{\varphi}$) are shown in Fig. \ref{fig2} (c),(d). Each fragment shows results for three values of the frequency $\omega_{ex}$: near the synchronization region (limit cycles 1 and 2) and inside the synchronization region (equilibrium point 3). In the case of Fig. \ref{fig2} (c), an equilibrium point arises at the limit cycle, which indicates the saddle-node infinite period (SNIPER) bifurcation characterizing the  phase-locking region boundary. It can be seen in Fig. \ref{fig2} (d) that passing through the 
synchronization boundary leads to contraction of limit cycles to a point. It indicates the Andronov--Hopf bifurcation, corresponding to suppression of oscillations by an external influence.

\begin{figure}[t!]
\centering
\parbox[c]{.49\linewidth}{
  \includegraphics[width=\linewidth]{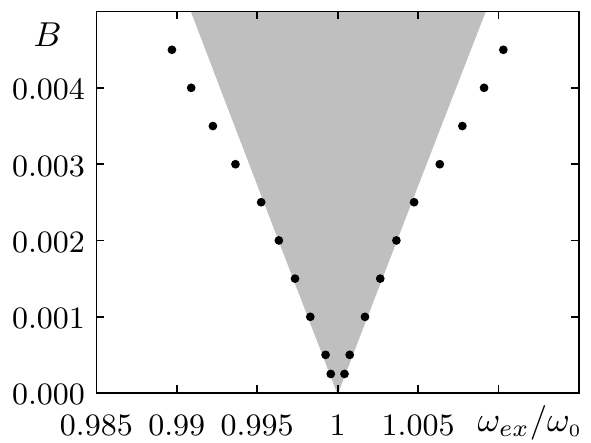}
}
\parbox[c]{.49\linewidth}{
  \includegraphics[width=\linewidth]{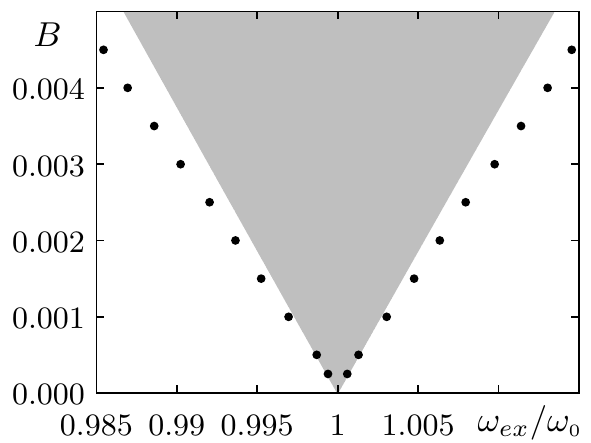}
}
\center (a)   \hspace{7cm}    (b)

\caption{Synchronization areas of the memristive oscillator for two initial values of the variable $z$: $z_{0} = 0$ (panel (a)) and $z_{0} = 0.1$ (panel (b)) with $x(0) = y(0) = 0$. The black circles correspond to the synchronization region boundaries obtained numerically for the system (\ref{model}), the toned region is the synchronization region in accordance with (\ref{cond}). The system parameters are $\alpha = 1$, $\beta = 0.035$, $\gamma = 1$, $a = 0.02$, $b = 0.8$.}
\label{fig3}
\end{figure}

\begin{figure}[t]
\centering
\parbox[c]{.49\linewidth}{
  \includegraphics[width=\linewidth]{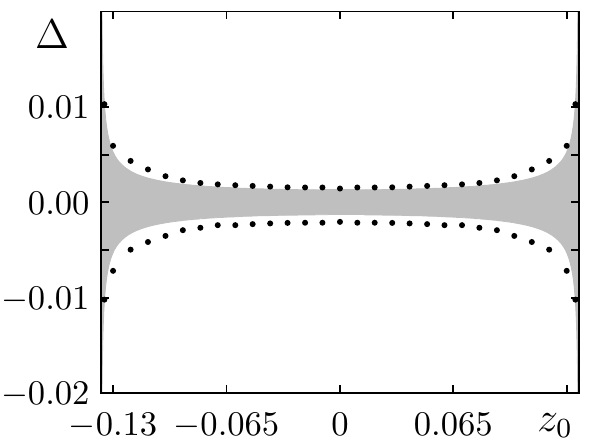}
}
\caption{Dependence of synchronization area boundaries on initial state $z_0$ on plane ($z_{0}$,$\Delta$) at fixed amplitude of driving impact, $B = 0.001$. The black circles correspond to the numerical study of the system (\ref{model}) (other initial conditions are $x(0) = y(0) = 0$). The toned area corresponds to the fulfilment of the conditions (\ref{cond}) for the phase model. Other parameters are: $\alpha = 1$, $\beta = 0.035$, $\gamma = 1$, $a = 0.02$, $b = 0.8$.}
\label{fig4}
\end{figure}

The results presented above show that the phase model describes well the forced synchronization of the memristive oscillator (\ref{model}) only at very small amplitudes of the influence. In such a way, at $z(0) = 0$ a good agreement is observed only for $B \leq 0.002$. However, the phase model reflects the principal feature of synchronization of the memristive oscillator: the dependence of the synchronization effect on the initial conditions (in particular, on the initial value of the variable controlling the memristor $z(0) = z_0$)\footnote{Generally speaking, the amplitude of autonomous oscillations, and hence, the synchronization boundary in the system (\ref{model}) depends not only on $z(0)$, but also on the initial value $x(0)$. However it was assumed during derivation of truncated equations (\ref{eq-rho-phi}) that $x(0) = 0$.}. Let us consider how the value of $z_0$ affects the synchronization mode. Figure \ref{fig3} shows synchronization areas estimated in a numerical simulation of the system (\ref{model}) and areas obtained by using the approximate expression (\ref{cond}) for two initial values of the variable $z$: $z_{0} = 0$ [Fig. \ref {fig3}(a)] and $z_{0} = 0.1$ [Fig. \ref{fig3} (b)] with $x(0)=y(0)=0$. One can see a significant difference in the width of the synchronization regions obtained in numerical study and derived using the phase approximation.
%
%%%%%%%%%%%%%%%%%%%%%%%%%%% Рис.3 %%%%%%%%%%%%%%%%%%%%%%%%%%%%%%%%%%%%%%%%%%%%%%%%%%%%
%
%%%%%%%%%%%%%%%%%%%%%%%%%%%%%%%%%%%%%%%%%%%%%%%%%%%%%%%%%%%%%%%%%%%%%%%%%%%%%%%%%%%%%%%%
%
It should be emphasized that the dependence of a steady-state synchronous or non-synchronous mode is not a consequence of the usual multistability (i.e. coexistence of several attractors in the phase space of the system (\ref{model})). The boundaries of the synchronization region are continuously shifted when the initial value $z_{0}$ changes. This fact follows from the approximate phase model. In order to confirm this effect, Fig. \ref{fig4} shows the dependence of the detuning parameter $\Delta$ interval corresponding to the  synchronization on the value $z_{0}$ obtained at $B = 0.001$, $x(0)~=~y(0)~=~0$. The results obtained for the original system (\ref{model}) are compared with the approximate analytical results obtained in accordance with (\ref{cond}). Since the amplitude of the impact is chosen to be small, there is a good agreement between the results of the numerical study of the initial system and the analytical estimation of the dependence in the phase approximation. The width of the synchronization region is minimal at $z_{0}=0$ and increases with increasing $|z_{0}|$ tending to infinity when some boundary values $z_{0} = \pm z_{m}$ are reached. Estimation based on the phase model (\ref{eq-phi}) gives the value
\begin{eqnarray}
\label{zm}
z_{m}=\sqrt{\frac{\beta-a}{b}} \approx 0.13693.
\end{eqnarray}
In case $|z_{0}| \geq z_{m}$ beats are not observed regardless the amplitude of the impact. This is obviously due to the fact that in this case there are no oscillations in the autonomous oscillator (\ref{model}) and trajectories tend to a stable equilibrium (one of points on the $0Z$ axis with the coordinate $|z| \geq z_{m}$).
%
%%%%%%%%%%%%%%%%%%%%%%%%%%% Fig.4 %%%%%%%%%%%%%%%%%%%%%%%%%%%%%%%%%%%%%%%%%%%%%%%%%%%%
%
%%%%%%%%%%%%%%%%%%%%%%%%%%%%%%%%%%%%%%%%%%%%%%%%%%%%%%%%%%%%%%%%%%%%%%%%%%%%%%%%%%%%%%%%
%

\section{Conclusions}
Both numerical and approximate analytical studies of the behavior of a memristive oscillator under an external harmonic influence show the presence of a forced synchronization in a finite range of the impact frequency values (and, accordingly, frequency detuning). Two synchronization mechanisms have been established: capture of the phase and  frequency of the oscillator and suppression of oscillations by an external signal. The effect of phase locking is observed at very small values of the amplitude of the impact, and in this case, the non-autonomous memristive oscillator is described quite well by an approximate phase equation. The presence of frequency-phase locking testifies self-oscillatory nature of the memristive oscillator dynamics. However, due to the fact that the characteristics of self-sustained oscillations (mainly amplitude) in a memristive self-oscillator continuously depend on the initial conditions (especially on the initial value of the variable $z$, which controls memristive conductivity) the synchronization phenomenon has its own peculiarities. The boundaries of the synchronization area are continuously shifted as the initial value of $z_0$ changes. Varying this values, one can observe both the synchronous periodic mode and the beat mode at the same parameters of the self-oscillator and external force. As our studies have shown, this dependence is also observed in the interaction of traditional self-sustained oscillators connected via a memristive element \cite{korneev2019} and, apparently, is a distinctive feature of synchronization in memristive systems.

\section*{Acknowledgments}
This work was supported by the Russian Ministry of Education and Science (project code 3.8616.2017/8.9).

%

%\bibliographystyle{epj}
%\bibliography{bibliography}

\end{document}